\documentclass[pre,twocolumn,preprintnumbers,aps,showpacs,amsmath,amssymb]{revtex4}

\usepackage{graphicx}
\usepackage{dcolumn}
\usepackage{bm}

\begin{document}

\title{Oscillatory Motion of a Camphor Grain in a One-Dimensional Finite Region}

\author{Yuki Koyano}
\affiliation{Department of Physics, Graduate School of Science, Chiba University, Chiba 263-8522, Japan}

\author{Tatsunari Sakurai}
\affiliation{Department of Physics, Graduate School of Science, Chiba University, Chiba 263-8522, Japan}

\author{Hiroyuki Kitahata\footnote{Corresponding author. E-mail: kitahata@chiba-u.jp.}}
\affiliation{Department of Physics, Graduate School of Science, Chiba University, Chiba 263-8522, Japan}

\begin{abstract}
The motion of a self-propelled particle is affected by its surroundings, such as boundaries or external fields.
In this paper, we investigated the bifurcation of the motion of a camphor grain, as a simple actual self-propelled system, confined in a one-dimensional finite region.
A camphor grain exhibits oscillatory motion or remains at rest around the center position in a one-dimensional finite water channel, depending on the length of the water channel and the resistance coefficient.
A mathematical model including the boundary effect is analytically reduced to an ordinary differential equation.
Linear stability analysis reveals that the Hopf bifurcation occurs, reflecting the symmetry of the system.
\end{abstract}


\maketitle

\section{Introduction}

Self-propelled motion under nonequilibrium conditions has been receiving increasing attention~\cite{Ramaswamy,Mikhailov,Vicsek}.
The motion of living beings is considered to be self-propelled, and it has often been investigated by using real living systems such as keratocytes, slime molds, locusts, and so on~\cite{Keren,Tanimoto,Taniguchi,Ariel}.
However, well-controlled experiments are difficult to conduct in these biological systems, since the experiments are affected by environmental fluctuations and differences between individual systems.
In addition, biological systems comprise complex physiological processes, rendering their motion difficult to understand based on physico-chemical elementary processes.
To address these disadvantages, physico-chemical systems are useful, as they are simple and easily controlled.
A camphor grain floating on water, which diffuses camphor molecules around the grain, is one of the most studied physico-chemical self-propelled particles~\cite{Skey,Rayleigh1890,Nakata2015,Nakata1997,Nakata2000}.
The camphor molecules act as surfactants and reduce the surface tension of the water; the camphor grain is then driven by the surface tension gradient around it.

In general, the symmetry of the self-propelled system directly results in the symmetry of the motion.
The system symmetry is characterized by various factors such as the shape of the particle~\cite{Paxton,Howse,Lowen,Hayakawa,Nakata1997,Nakata2000,OhtaPRL,Kitahata}, the external field~\cite{Young,Cejkova,Diguet}, and the system boundaries~\cite{Linke,SuminoPRL,SuminoPRE,Nishi,Hayashima}.
In an asymmetric system, motion reflecting the embedded asymmetry is observed~\cite{Paxton,Howse,Lowen,Hayakawa,Nakata1997,Nakata2000,Young,Cejkova,Diguet,Linke}.
Typical examples of the asymmetric systems are the systems that exhibit taxis under a certain gradient~\cite{Young,Cejkova,Diguet}.
On the other hand, in a symmetric system, the motion emerges through spontaneous symmetry breaking~\cite{Domingues,Cira,Nagai,Pimienta,OhtaPRL,Kitahata,Koyano,SuminoPRL,SuminoPRE,Nishi,Nagayama,Hayashima,Ohta2009}.
In such systems, when the rest state becomes unstable, the motion appears to reflect the system symmetry, e.g., translation, oscillation, and rotation, in isotropic, mirror-symmetric, and axisymmetric systems, respectively~\cite{Koyano}.

A symmetric camphor grain, i.e., a disk-shaped camphor grain, in a symmetric water chamber is classified into the latter case.
When a camphor grain is located at the center position of a water chamber, it can stop because the force originating from the surface tension can be balanced owing to the symmetric profile of camphor molecules.
However, a camphor grain at rest can be disturbed by the fluctuation.
Once the camphor grain begins to move, the asymmetry of the concentration field can be enhanced by the motion.
In such cases, the rest state is unstable, and the camphor grain continues moving.

In previous studies, the switching of the rest state and the stable motion in one-dimensional infinite and finite systems were investigated.
In a one-dimensional infinite system, a camphor grain either rests at an arbitrary position or translates with a constant velocity depending on the resistance coefficient of the camphor grain~\cite{Nagayama}.
In terms of the bifurcation theory, this was understood as pitchfork bifurcation.
Recently, Nishi {\it et al.} investigated mode bifurcation in the motion of single or multiple camphor grains in a one-dimensional system with periodic boundary conditions~\cite{Nishi}.
As for one-dimensional finite regions with inversion symmetry, Hayashima {\it et al.} reported that oscillatory motion was observed in the experiments~\cite{Hayashima}.
However, the bifurcation structure has not been investigated in detail.

In the present paper, we focus on the spontaneous symmetry breaking in the motion of a self-propelled particle confined in a one-dimensional finite system with inversion symmetry.
We aim to understand the motion in such systems in terms of bifurcation theory by considering a real physico-chemical system; here, we adopt the camphor-water system.
We experimentally and theoretically investigate the motion of a camphor grain in a one-dimensional finite system, depending on the system size and the viscosity of the aqueous phase.

In the next section, the experimental results of the motion of a camphor grain in a one-dimensional water channel were introduced.
Section~\ref{III} describes the framework of the mathematical model for the motion of a camphor grain.
In Section~\ref{IV}, the model was analytically reduced into an ordinary differential equation, and the bifurcation structure was investigated by a theoretical analysis.
We also confirmed the analytical results by performing numerical calculations in Section~\ref{V}.
This paper concludes with the discussion on the several unique features of a finite system.

\section{Experiments\label{II}}

\begin{figure}
	\begin{center}
		\includegraphics{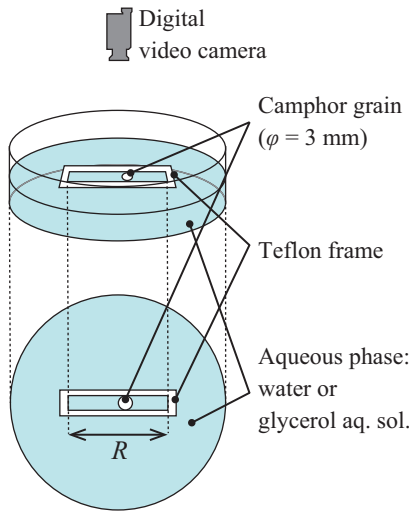}
		\caption{Schematic illustration of the experimental setup.}
		\label{exp_setup}
	\end{center}
\end{figure}

Figure~\ref{exp_setup} shows the schematic illustration of the experimental setup.
A glycerol aqueous solution was prepared by mixing glycerol (Wako, Japan) with pure water, which was prepared with the Millipore water purifying system (UV3, Merck, Germany).
The water channel was made of a Teflon sheet, whose thickness was 1~mm, with a rectangular aperture (width: 4~mm, length: $R$, where $R$ = 15, 20, 25, 30, 35, 40, 45, or 50~mm).
The channel was fixed at the surface of pure water or glycerol aqueous solution in a Petri dish (inner diameter: 200~mm, volume of the aqueous phase: 250~mL).
A camphor grain was prepared by pressing camphor (Wako, Japan) with a pill maker (Kyoto Pastec, Japan).
The camphor grain has a discoid shape with a diameter of 3~mm and a thickness of {\it ca.} 1~mm.
The camphor grain was put into the water channel.
The motion of the camphor grain was captured for 6~min at 30~frames per second (fps) with a digital video camera (iVIS HV30, Canon, Japan).

By changing the concentration of glycerol in the aqueous phase, the friction exerted on a camphor grain was changed.
The viscosity of the glycerol aqueous solution was measured by a vibrational viscometer (SV-10A, A\&D, Japan).
All the experiments were conducted at room temperature.

The obtained videos were analyzed using an image-processing software (ImageJ, NIH, USA).
For the data analysis, we used the videos from 1 to 6~min after a camphor grain was floated.
The reason for using videos from 1~min was that the motion of the camphor grain converged to a stable oscillation after several oscillations, and the characteristic time scale of the oscillation of the camphor grain was of about 1-2~s.
Additionally, the reason for using videos up to 6~min was that regular oscillations of the camphor grain were disturbed at around 10~min after it was floated.

\begin{figure}
	\begin{center}
		\includegraphics{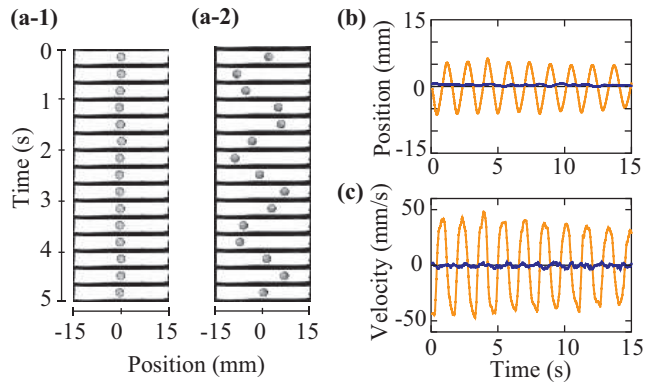}
		\caption{Experimental results.
		(a) Snapshots of the camphor grain at the surface of (a-1) 3.5~mol/L and (a-2) 2.5~mol/L glycerol aqueous solution.
		The viscosities of the aqueous phases were (a-1) 3.1~mPa$\cdot$s and (a-2) 2.3~mPa$\cdot$s. The length of the channel was 30~mm in both cases.
		(b) Time evolution of the position.
		(c) Time evolution of the velocity.
		The plots colored with blue (dark gray) and orange (light gray) correspond to (a-1) and (a-2), respectively.}
		\label{exp_sys_snap_xt_vt}
	\end{center}
\end{figure}

In the experiments, we observed rest (Fig.~\ref{exp_sys_snap_xt_vt}(a-1)) or oscillatory states (Fig.~\ref{exp_sys_snap_xt_vt}(a-2)) around the center position of the water channel depending on the length of the water channel and the viscosity of the aqueous phase.

\begin{figure}
	\begin{center}
		\includegraphics{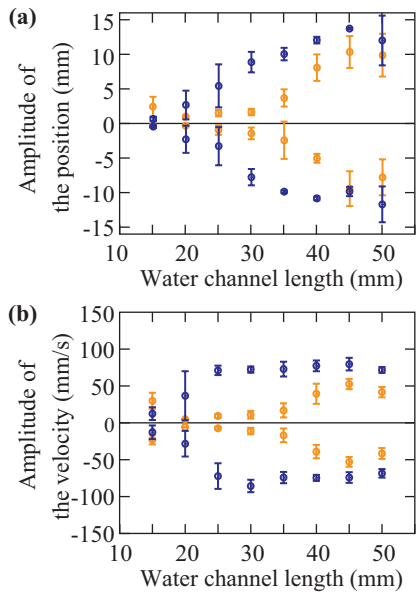}
		\caption{Dependence of the motion of the camphor grain on the water channel length.
		(a) Amplitude of the position.
		(b) Amplitude of the velocity.
		The results for pure water (blue (dark gray), 1.0 mPa$\cdot$s) and 3.5~mol/L glycerol aqueous solution (orange (light gray), 3.1 mPa$\cdot$s) are shown.
		Error bars show the standard deviation.}
		\label{exp_amp_vamp}
	\end{center}
\end{figure}

From the time series of the position of a camphor grain, the averaged amplitudes of position and velocity were obtained as shown in Fig.~\ref{exp_amp_vamp}.
The values were almost zero for shorter water channels, while they had finite values for longer water channels.
The threshold value of water channel length between these two situations were $\sim$20~mm for pure water and $\sim$35~mm for the 3.5~mol/L glycerol aqueous solution.
For the systems with sizes above the threshold values, the amplitudes of the position and velocity were greater for smaller viscosity.
The detailed analysis method is presented in Section~\ref{App1} of the Appendix.

\begin{figure}
	\begin{center}
		\includegraphics{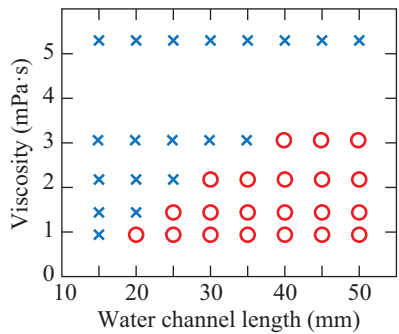}
		\caption{(Color online) Phase diagram obtained by experiments.
		Circles and crosses correspond to the oscillatory and rest state, respectively.
		The concentrations of the glycerol aqueous solutions were 0~mol/L (pure water) for 1.0 mPa$\cdot$s, 1.0~mol/L for 1.3 mPa$\cdot$s, 2.5~mol/L for 2.3 mPa$\cdot$s, 3.5~mol/L for 3.1 mPa$\cdot$s, and 5.0~mol/L for 5.3 mPa$\cdot$s.}
		\label{exp_pd}
	\end{center}
\end{figure}

The behaviors of the camphor grain were classified into being in oscillatory and rest states, and summarized as a phase diagram in Fig.~\ref{exp_pd}.
The oscillatory and rest states were defined as the states in which amplitudes of averaged velocity were greater and smaller than 30~mm/s, respectively.
It was clearly observed that the oscillatory state was realized for longer water channels or smaller viscosity of the aqueous phase.

\section{Mathematical Model\label{III}}

The model for the motion of a camphor grain is constructed based on the previous work by Nagayama {\it et al.}~\cite{Nagayama}.
We consider the motion of a camphor grain in a one-dimensional finite system with the size $R$.
The position of the camphor grain is represented by $X(t) \in [0,R]$, and the concentration field of camphor molecules at the water surface is denoted as $c(x,t)$, where $x \in [0,R]$ and $t > 0$ are the position and time, respectively.
The mathematical model is composed of the evolutional equations of $X(t)$ and $c(x,t)$.
In the present paper, a camphor grain is regarded as a point particle, which is different from previous papers~\cite{Nagayama,Nishi,Kitahata}.

The equation of motion for the camphor grain is
\begin{equation}
\label{eq.of.m.}
m \frac{d^2 X}{dt^2} = - \eta \frac{d X}{dt} + F,
\end{equation}
where $F = F(X(t);c)$ denotes the driving force originating from the surface tension gradient around the camphor grain.
$m$ and $\eta$ represent the mass and resistance coefficient of the camphor grain, respectively.
Here, $m$ is regarded as a constant, since the dissolution of the camphor molecule is sufficiently slower than the characteristic time scale of the camphor grain motion.

The surface tension, $\gamma$, depends on the surface concentration of camphor molecules, $c$, at each position.
Here, we assume that $\gamma$ is a linear decreasing function; $\gamma(c) = -\Gamma c + \gamma_0$, where $\gamma_0$ is the surface tension of pure water and $\Gamma$ is a positive constant.
The driving force by the surface tension is defined as 
\begin{align}
F(X(t);c) = k \left [ \gamma(c(X(t)+\epsilon)) - \gamma(c(X(t)-\epsilon)) \right ], \label{def_driving_force}
\end{align}
where $k$ is a positive constant, and $\epsilon$ is an infinitesimally small positive parameter.
Equation~\eqref{def_driving_force} shows that the driving force is proportional to the difference in the surface tension between the right and left sides of the camphor grain.
By expanding it around $\epsilon = 0$, we obtain
\begin{align}
F(X(t);c) = -K \left ( \left . \frac{\partial c}{\partial x}\right |_{x=X(t)+0} + \left . \frac{\partial c}{\partial x} \right |_{x=X(t)-0} \right ), \label{driving_force}
\end{align}
where $K = k \epsilon \Gamma > 0$.
We assume that $K = k \epsilon \Gamma$ is finite even when $\epsilon \to +0$.

It is known that camphor molecules sublimate into air.
The time evolution of $c$, the concentration field at the water surface, is composed of the diffusion term, the sublimation term, and the source term as
\begin{equation}
\label{d.eq.}
\frac{\partial c}{\partial t} = D \frac{\partial ^2 c}{\partial x^2} - \alpha c + f,
\end{equation}
where $D$ is the diffusion coefficient of the camphor molecules at the water surface, and $\alpha$ is their sublimation rate from the water surface.
$f$ represents the supply of the camphor molecules from the camphor grain,
\begin{equation}
f(x;X(t)) = c_0 \delta (x-X(t)), \label{source}
\end{equation}
where $\delta(x)$ is the Dirac delta function reflecting that the camphor grain is regarded as a point particle.
Equation~\eqref{source} indicates that the camphor molecules are dissolved by $c_0$ per unit time at $x=X(t)$.
The Neumann boundary condition for Eq.~\eqref{d.eq.} is imposed:
\begin{equation}
\label{n.c.}
\left . \frac{\partial c}{\partial x} \right |_{x=0,R} = 0,
\end{equation}
which represents that there is no diffusion flux at the boundaries from Fick's law of diffusion.

We derive the dimensionless equations:
\begin{gather}
\label{nd_motion_eq.} \frac{d^2 \tilde{X}}{d\tilde{t}^2} = - \tilde{\eta} \frac{d \tilde{X}}{d\tilde{t}} + \tilde{F}(\tilde{X};\tilde{c}),\\
\label{nd_d.f.} \tilde{F}(\tilde{X};\tilde{c}) = - \left( \left . \frac{\partial \tilde{c}}{\partial \tilde{x}}\right |_{\tilde{x}=\tilde{X}+0} + \left . \frac{\partial \tilde{c}}{\partial \tilde{x}} \right |_{\tilde{x}=\tilde{X}-0} \right),\\
\label{nd_d.eq.} \frac{\partial \tilde{c}}{\partial \tilde{t}} = \frac{\partial^2 \tilde{c}}{\partial \tilde{x}^2} - \tilde{c} + \tilde{f}(\tilde{x};\tilde{X}), \\
\tilde{f}(\tilde{x};\tilde{X}) = \delta (\tilde{x}-\tilde{X}),
\end{gather}
where $\tilde{x} = \sqrt{\alpha/D} x$, $\tilde{t}=\alpha t$, $\tilde{X}(\tilde{t}) = \sqrt{\alpha / D} X(\tilde{t} / \alpha)$, $\tilde{c} = c / c_0$, $\tilde{m} = m \alpha D / (K c_0)$, and $\tilde{\eta} = \eta D / (K c_0)$.
Hereafter, we use these dimensionless equations with the tildes omitted.
The detailed method of nondimensionalization is presented in Section~\ref{App2} in the Appendix.

\section{Reduction and analysis of the model\label{IV}}

In our mathematical model, the rest state at the system center should exist from the viewpoint of the symmetric property.
The experimental results in Fig.~\ref{exp_amp_vamp} suggest that the oscillatory state emerges through the destabilization of the rest state.
Thus, we analytically reduced the model equations to an ordinary differential equation around the rest state and investigated its linear stability.

First, $c(x,t)$ and $f(x;X(t))$ are expanded as the Fourier series with the basis, $\cos(k\pi x/R)$ ($k = 0, 1, 2, \cdots$), which satisfies the Neumann condition,
\begin{gather}
c(x,t) = \frac{1}{R} \sum_{k=0}^{\infty} c_k (t) \cos \left(\frac{k\pi x}{R} \right), \\
f(x;X(t)) = \frac{1}{R} \sum_{k=0}^{\infty} f_k (X(t)) \cos \left(\frac{k\pi x}{R} \right).
\end{gather} 
$f_k(X(t))$ is calculated as $f_0 (X(t)) = 1$ and $f_k(X(t)) = 2 \cos (k \pi X(t) / R )$ for $k \geq 1$.
The equation for the concentration field in the wave-number space is 
\begin{eqnarray}
\frac{d c_k}{d t} = \left ( i \frac{k \pi}{R} \right )^2 c_k - c_k +  f_k(X(t)).
\end{eqnarray}
We construct $c_k$ by using the Green's function.
The Green's function $g_k(t)$ is defined as
\begin{eqnarray}
\left ( \frac{d}{d t} + \kappa^2  + 1 \right ) g_k(t) &=& \delta(t),
\end{eqnarray}
where $\kappa = \kappa(k) = k \pi/R$, and it was calculated as
\begin{align}
g_k (t) = 
\left \{
\begin{array}{ll}
0 & (t < 0),\\
\displaystyle{\exp \left(-\left ( \kappa^2 + 1 \right ) t \right )} & (t > 0).
\end{array}
\right .
\end{align}
Thus, $c_k(t)$ is described in an integral form:
\begin{align}
c_k(t) = \int_{-\infty}^{t} d{t}' f_{k} \left( X({t}') \right ) \exp \left(-\left ( \kappa^2 + 1 \right ) (t - {t}')\right ). \label{integralform}
\end{align}
By expanding Eq.~\eqref{integralform} using partial integration,
$c_k$ is expanded with regard to the position $X$, velocity $\dot{X}$, acceleration $\ddot{X}$, and so on, of the camphor grain~\cite{Ohta2009}, where dots ($\,\dot{}\,$) denote the time derivative.
By converting the expanded $c_k$ using the inverse Fourier transform, we obtain the concentration field in real space as follows:
\begin{align}
c(x,t) = & c_0 (x,X) \nonumber \\
& + \dot{X} c_1 (x,X) + \dot{X}^2 c_2 (x,X) + \dot{X}^3 c_3 (x,X) + \cdots \nonumber \\
& + \ddot{X} c_4 (x,X) + \cdots \nonumber \\
& + \mathrm{(higher \; order \; terms \; \& \; cross \; terms)}. \label{exp.cnc.}
\end{align}
The driving force is calculated using Eq.~\eqref{nd_d.f.} with Eq.~(\ref{exp.cnc.}) as a function of $X$, $\dot{X}$, and $\ddot{X}$.

To investigate the linear stability of the rest state at the system center, we expand the driving force, Eq.~\eqref{nd_d.f.}, with Eq.~(\ref{exp.cnc.}) around $X=R/2$.
Here, we set the shift from the center position as $\delta X = X - R/2$.
We then obtain the second-order ordinary differential equation for $\delta X$:
\begin{eqnarray}
m \ddot{\delta X} = - \eta \dot{\delta X} + F(\delta X, \dot{\delta X}, \ddot{\delta X}),
\label{rdc.eq}
\end{eqnarray}
where
\begin{align}
&F(\delta X, \dot{\delta X}, \ddot{\delta X}) \nonumber \\
=& A(R) \delta X + B(R) (\delta X)^3 + C(R) \dot{\delta X} + E(R) (\delta X)^2 \dot{\delta X} \nonumber \\
& + G(R) \ddot{\delta X} + H(R) \delta X  \left ( \dot{\delta X}\right ) ^2 + I(R) \left ( \dot{\delta X} \right ) ^3.
\label{df}
\end{align}
The coefficients of each term of the driving force, $A(R)$, $B(R)$, $C(R)$, $E(R)$, $G(R)$, $H(R)$, and $I(R)$ are functions of the system size, $R$.
Their explicit forms are obtained as
\begin{align}
A(R) =& - \frac{2}{\sinh R},\\
B(R) =& - \frac{4}{3\sinh R}, \\
C(R) =& \frac{(\cosh R -1)(\sinh R + R)}{2 \sinh^2 R}, \\
E(R) =& \frac{- 3 \sinh R + R \cosh R}{\sinh^2 R}, \\
G(R) =& - \frac{1}{8 \sinh^3 R} (\cosh R -1) \nonumber\\
& \times \left [ \sinh R (\sinh R - R) + R^2(\cosh R - 1) \right ], \\
H(R) =& - \frac{1}{4 \sinh^3 R} \nonumber \\
& \times \left [\sinh R (3 \sinh R - 5 R \cosh R) \right . \nonumber \\
& \qquad \qquad \left . + R^2 (2 + \sinh^2 R) \right ], \\
I(R) =& - \frac{1}{48 \sinh^4 R} (\cosh R -1)^2 \nonumber \\
& \times \left [ R^3 (2 - \cosh R) + 6 R^2 \sinh R \right . \nonumber \\
& \qquad \qquad \left . + 3 (\cosh R + 1)(\sinh R - R) \right ].
\end{align}
It is noted that $C(R)$ is positive, and $A(R)$, $B(R)$, $G(R)$, and $I(R)$ are negative for $R>0$.
The details of this calculation are provided in Section~\ref{App3} of the Appendix.

\begin{figure}
	\begin{center}
		\includegraphics{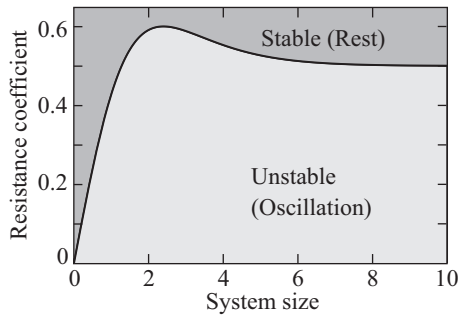}
		\caption{Bifurcation diagram obtained by the theoretical analysis. The curve describes $\eta=C(R)$.
		In the dark and light gray regions, the fixed point is stable and unstable, respectively.
		Here, the fixed point corresponds to the rest state of the camphor grain at the system center.
		The oscillatory motion of the camphor grain around the system center can be observed in experiments when the fixed point is unstable.
		}
		\label{pd_ac}
	\end{center}
\end{figure}

It is confirmed that $(\delta X, \dot{\delta X}) = (0,0)$ is a fixed point that corresponds to the situation where the camphor grain stops at the center position of the system.
Linear stability analysis reveals that the fixed point becomes unstable for $\eta < C(R)$ through Hopf bifurcation, as shown in Fig.~\ref{pd_ac}.
The unstable fixed point implies the existence of a stable limit cycle, which corresponds to the oscillatory motion of the camphor grain around the system center.
As shown in Fig.~\ref{pd_ac}, the bifurcation curve $C(R)$ has a peak around $R = 2$. 
This peak indicates that a camphor grain can continue moving with the largest resistance coefficient in the water channel with the length of $R \sim 2$.
The reason for the peak is shown in the Section~\ref{App4} of the Appendix.
However, it is reported that the existence of the peak depends on the relation between the concentration and surface tension~\cite{Nishi}. Weakly nonlinear analysis shows that the Hopf bifurcation is supercritical for $3I(R)\omega^2+E(R)>0$ and subcritical for $3I(R)\omega^2+E(R)<0$, where $\omega = \sqrt{-A(R)/(m-G(R))}$~\cite{Guckenheimer}. 
(See Section~\ref{App5} in the Appendix for more information on the detailed bifurcation structure.)

\section{Numerical Calculation\label{V}}

For comparison with the results obtained by the theoretical analysis, we performed numerical calculations based on Eqs.~\eqref{nd_motion_eq.} and \eqref{nd_d.eq.}.
We calculated Eq.~\eqref{nd_motion_eq.} with the Euler method and Eq.~\eqref{nd_d.eq.} with the implicit method.
The spatial mesh and the time step was set to $10^{-3}$ and $10^{-5}$, respectively.

\begin{figure}
	\begin{center}
		\includegraphics{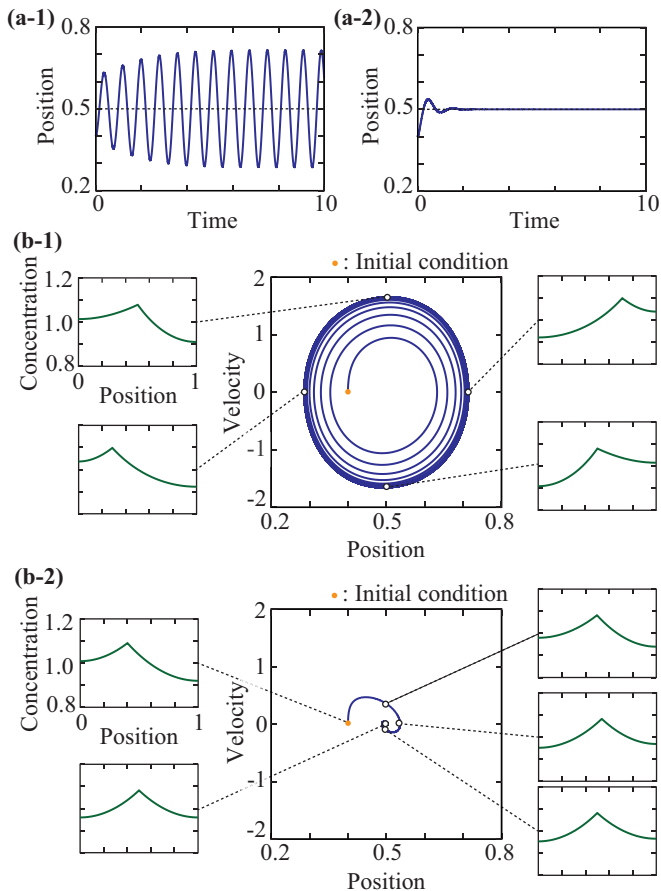}
		\caption{Representative numerical results.
		(a) Time series of the camphor grain position $X$. (b) Trajectories on the position-velocity ($X$-$\dot{X}$) phase space, and the concentration field, $c$, at each time.
		(a-1,b-1) Oscillatory motion.
		(a-2,b-2) Damping oscillation to the rest state.
		The system size was $R=1$ in both cases and the resistance coefficient, $\eta$, was set to (a-1,b-1) $\eta=0.3$ and (a-2,b-2) $\eta=0.5$.}
		\label{nc_orbits}
	\end{center}
\end{figure}

The representative numerical results are shown in Fig.~\ref{nc_orbits}.
The camphor grain exhibited a limit-cycle or damping oscillation depending on the resistance coefficient, $\eta$.
For smaller resistance coefficients, the camphor grain exhibited a limit-cycle oscillation, and the concentration field also changed periodically following the motion of the camphor grain, as shown in Fig.~\ref{nc_orbits}(a-1,b-1).
In contrast, for greater resistance coefficients, the camphor grain approached the center position of the system.
Eventually, the grain stopped at the center position, and the profile of the concentration field became symmetric, as shown in Fig.~\ref{nc_orbits}(a-2,b-2).

\begin{figure}
	\begin{center}
		\includegraphics{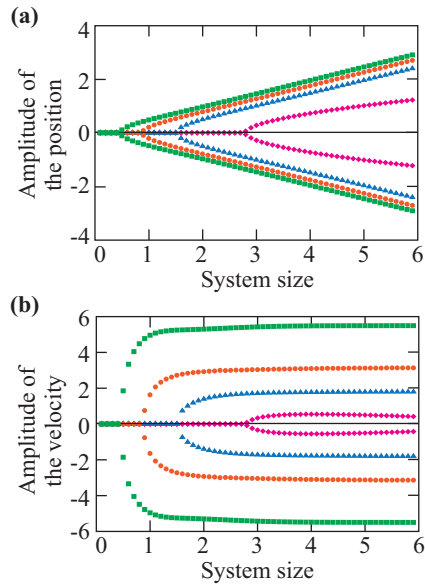}
		\caption{Dependence of the camphor grain motion on the system size, $R$.
		The amplitudes of (a) the position and (b) the velocity are plotted against the system size, $R$, for each resistance coefficient, $\eta$.
		The green squares, orange circles, blue triangles, and pink diamonds correspond to $\eta=$ 0.2, 0.3, 0.4, and 0.5, respectively.}
		\label{nc_amp_vamp}
	\end{center}
\end{figure}

To investigate the threshold between the oscillatory and rest states, we calculated trajectories until they sufficiently converged to a limit-cycle or fixed point.
We then evaluated the amplitudes of the position and velocity, and plotted them against the system size, $R$, for each resistance coefficient, $\eta$, as shown in Fig.~\ref{nc_amp_vamp}.
The amplitudes of the position and velocity were zero below the thresholds and they grew above the thresholds.
These characteristic behaviors implied the Hopf bifurcation.
The threshold of the system size between the oscillatory and rest states shifted to be greater for greater resistance coefficients, and the amplitudes of the position and velocity were greater for smaller resistance coefficient.

\begin{figure}
	\begin{center}
		\includegraphics{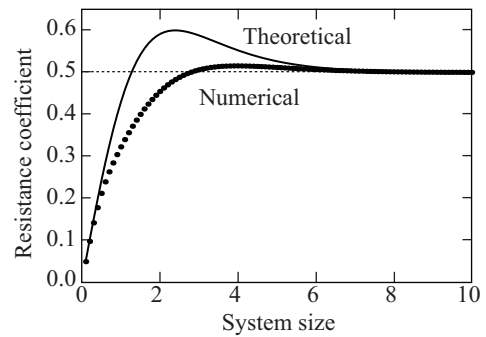}
		\caption{Comparison of the phase diagrams obtained numerically and analytically.
		Circles show the bifurcation points obtained numerically and the solid curve shows the bifurcation curve, $C(R)$, obtained analytically.}
		\label{pd_nc}
	\end{center}
\end{figure}

The threshold values are compared with the bifurcation points obtained analytically, as shown in Fig.~\ref{pd_nc}.
The numerical results roughly correspond to the analytical results; the value of the resistance coefficient at the bifurcation point is smaller for smaller system sizes and saturates to a certain value for greater system sizes.
The gap between the numerical and analytical results appears to be caused by factors such as the numerical error (mainly in the discretization of the Dirac delta function) and the truncation of the higher-order terms in the reduction process.

\begin{figure}
	\begin{center}
		\includegraphics{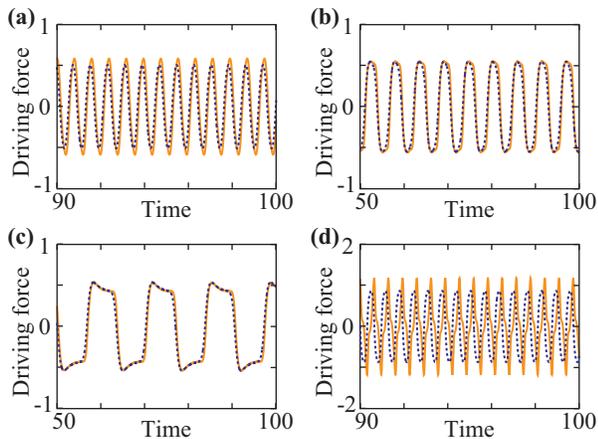}
		\caption{Time evolution of the driving force obtained with the original mathematical model (blue (dark gray) broken curve), and the reduced driving force (orange (light gray) solid curve).
		The parameters were (a) $R=1$ and $\eta=0.3$, (b) $R=4$ and $\eta=0.45$, (c) $R=8$ and $\eta=0.45$, and (d) $R=1$ and $\eta=0.25$.
		}
		\label{nc_force}
	\end{center}
\end{figure}

To evaluate the validity of the reduction of the mathematical model, we numerically compared the driving force calculated using the original model in Eqs.~\eqref{nd_motion_eq.} and \eqref{nd_d.eq.} with that calculated using the reduced driving force in Eq.~\eqref{df}, as shown in Fig.~\ref{nc_force}.
The time series of the reduced driving force was obtained by substituting the values of position, velocity, and acceleration numerically obtained with the original mathematical model.
For the parameter sets $(R,\eta)=(1,0.3)$, $(4,0.45)$, and $(8,0.45)$, the time series of the reduced driving force were similar to those directly obtained with the original mathematical model, as shown in Fig.~\ref{nc_force}(a-c).
On the while, the time series of the reduced driving force was significantly different from that obtained from the original mathematical model for the parameter set $(R,\eta)=(1,0.25)$, as shown in Fig.~\ref{nc_force}(d), since the parameter set was farther from the bifurcation point.
Therefore, the present reduction is acceptable for the situation near the bifurcation point in a semiquantitative manner.

\section{Discussion\label{VI}}

To confirm the validity of the theoretical analysis, we performed order estimation on the bifurcation points with regard to the dimensionless system size, $R$, and the dimensionless resistance coefficient, $\eta$.

First, the order of the dimensionless system size, $R$, was estimated.
The order of the water channel length in the experiments was $\sim$ 10~mm.
The diffusion length, $\sqrt{D / \alpha}$, was estimated as 10~mm from the distance between the boundary and the position at which the camphor grain was reflected.
Therefore, the order of $R$ is $R \sim$ 10~mm / 10~mm = 1.

Second, we consider the order of the dimensionless resistance coefficient, $\eta$.
The resistance coefficient was assumed to obey the Stokes' law, i.e., it is estimated to be $6 \pi \mu a$.
Here, $\mu$ is the viscosity of the aqueous phase and $a$ is the radius of the camphor grain, where $a=$ 1.5~mm in the experiments.
The typical value of $\mu$ was $\sim$ 1~mPa$\cdot$s for pure water and $\sim$ 10~mPa$\cdot$s for glycerol aqueous solution.
The driving force is proportional to the gradient of the concentration field in the neighborhood of the camphor grain.
The driving force is estimated to be $K c_0 / D$, which is obtained by the gradient of the concentration field for the rest state, $c_0 \exp(-\sqrt{\alpha/D}|x|) / (2\sqrt{\alpha D})$.
In the previous work, the driving force and sublimation rate were experimentally measured as {\it ca.} 1~$\mu$N and $(1.8 \pm 0.4) \times 10^{-2}$~$\mathrm{s}^{-1}$, respectively~\cite{Suematsu}.
Therefore,
\begin{align}
\eta = \frac{D}{K c_0} \sqrt{\frac{D}{\alpha}} \alpha \times 6 \pi \mu a \sim 6 \mu \; [\mathrm{Pa} \cdot \mathrm{s}]^{-1},
\end{align}
and we obtained $\eta \sim 10^{-2}$ for water and $\eta \sim 10^{-1}$ for glycerol aqueous solution.
Thus, the bifurcation point obtained analytically agrees with that obtained experimentally.
It is noted that the effective diffusion coefficient obtained experimentally in~\cite{Suematsu} is not adopted owing to the difference in spatial dimensionality.

\begin{figure}
	\begin{center}
		\includegraphics{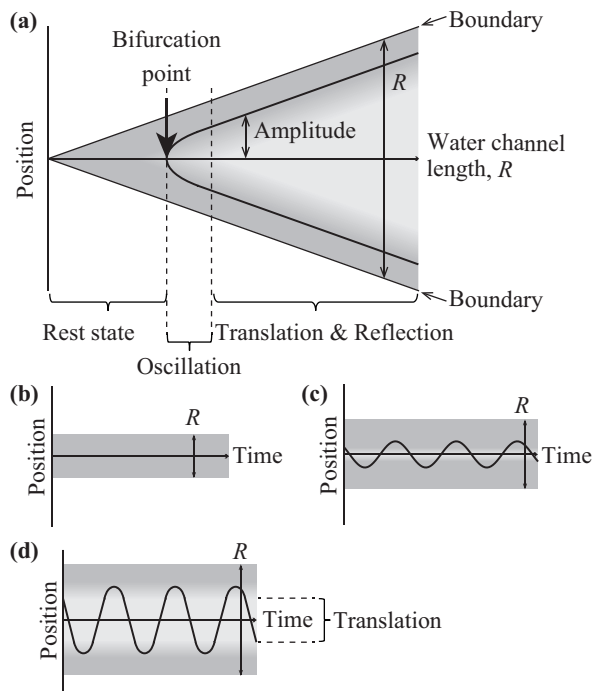}
		\caption{Schematic illustration on the behaviors of the camphor grain depending on the length of water channels with a constant viscosity.
		(a) Schematic bifurcation diagram showing the relation between the amplitude of motion and the system boundary.
		The darker-colored areas indicate the regions suffering from the effect of the boundaries, i.e., within the diffusion length from the boundaries.
		(b-d) Schematic description of the time series of the typical three motions depending on the system size.
		(b) Rest state.
		(c) Oscillation.
		(d) Translation and reflection.
		It is noted that there is no definite classification between (c) and (d).
		}
		\label{exp_sch_ill}
	\end{center}
\end{figure}

Here, we discuss the physical meaning of the bifurcation structure in the motion of the camphor grain in a one-dimensional finite region, as schematically shown in Fig.~\ref{exp_sch_ill}.
For a small system size, the camphor grain does not move since the camphor molecules are quickly saturated at the water surface and do not produce sufficient driving force.
Thus, the camphor grain stops at the center position as in Fig.~\ref{exp_sch_ill}(b), where the driving force balances.
By increasing the system size, the saturation of the camphor molecules becomes slower and the camphor grain begins to move.
The grain does not exhibit translational motion owing to the confinement of the boundaries, but it exhibits oscillation around the system center, as shown in Fig.~\ref{exp_sch_ill}(c).
For the greater system size, the amplitude of the position increases almost constantly and the amplitude of the velocity is saturated, as shown in Figs.~\ref{exp_amp_vamp} and \ref{nc_amp_vamp}
This behavior can be understood by considering the effect of the boundaries, which affect the motion of a camphor grain through the concentration field.
The characteristic length of the effect of the boundaries is considered to be the diffusion length of the concentration field.
For system sizes greater than the diffusion length, the effect of the boundaries is negligible except for their neighborhood.
Therefore, a camphor grain exhibits translational motion with an almost constant velocity that is determined only by the viscosity of the aqueous phase, and it is reflected by a boundary when the camphor grain is within the distance of the diffusion length from the boundary, as shown in Fig.~\ref{exp_sch_ill}(d).

To consider the translation in the region far from the boundaries as shown in Fig.~\ref{exp_sch_ill}(d), we discuss the motion of a camphor grain in an infinite region.
Equations~\eqref{nd_motion_eq.} and \eqref{nd_d.eq.} are reduced to the ordinary differential equation in the same way.
The difference in the model between the finite and infinite systems is only the boundary condition for the concentration field.
As a result of the reduction, the driving force $F$ is calculated as
\begin{eqnarray}
F =  \frac{1}{2} \dot{X}(t) - \frac{1}{8} \ddot{X}(t) - \frac{1}{16} (\dot{X}(t))^3 \label{df_ifnt},
\end{eqnarray}
where the higher-order terms are neglected.
It is noted that the driving force depends on the velocity and acceleration of the camphor grain but not on the position, since the infinite system has translational symmetry.
From linear stability analysis of the obtained ordinary differential equation, we confirm that pitchfork bifurcation occurs, which is consistent with the previous study~\cite{Nagayama}.
The translational velocity in the region far from the boundaries in the finite system is almost the same as that in the infinite system, since Eq.~\eqref{df} corresponds to Eq.~\eqref{df_ifnt} when $R \to \infty$.
Therefore, the motion in Fig.~\ref{exp_sch_ill}(d) is regarded as the combination of translation in an infinite system and reflection by the boundaries~\cite{Mimura,Chen}.

\section{Summary}

In the present paper, we investigated the motion of a self-propelled particle confined in a one-dimensional finite system with inversion symmetry by considering a camphor grain on water as a real system.
In the experiments, the oscillatory and rest states around the center of the water channel were observed depending on the length of the water channel and the viscosity of the aqueous phase.
The mathematical model including the time evolution of the position of the camphor grain and the profile of the camphor molecules was analytically reduced, and its bifurcation structure was theoretically analyzed.
The reduced model exhibits the Hopf bifurcation, which is consistent with the experimental results.
The analytical results were also confirmed by numerical calculations.

As an extension of the present work, it is interesting to consider the motion of a self-propelled particle in other systems with various symmetries such as a two-dimensional axisymmetric system, a two-dimensional square system, and a spherical system.
Especially in a two-dimensional axisymmetric system, a self-propelled particle can exhibit oscillatory or rotational motion, and the bifurcation structure can be more complex~\cite{Koyano}.
To understand such bifurcation structures based on real physico-chemical systems is a subject for future work.

\begin{acknowledgments}

We acknowledge Nobuhiko J. Suematsu, Masaharu Nagayama, Alexander S. Mikhailov, and Takao Ohta for valuable discussions.
This work was supported by JSPS KAKENHI Grant Numbers JP25103008 and JP15K05199.
This work was also supported in part by the Core-to-Core Program ``Nonequilibrium dynamics of soft matter and information'' to Y.K. and H.K., the Sasakawa Scientific Research Grant from the Japan Science Society to Y.K. (No.~28-225), and the Cooperative Research Program of ``Network Joint Research Center for Materials and Devices'' to Y.K. and H.K.

\end{acknowledgments}

\appendix

\section{Method for detecting in motion of a camphor grain\label{App1}}

\begin{figure}
	\begin{center}
		\includegraphics{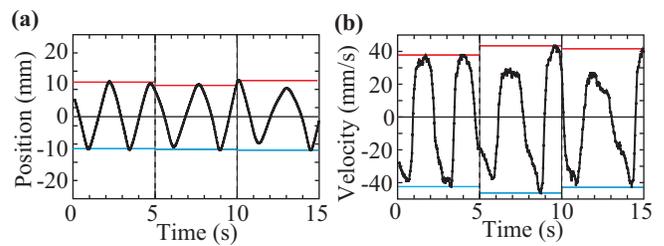}
		\caption{Schematic description of the method used to obtain the local maximum and minimum values from the time series of position and velocity.
		The plotted time series of (a) position and (b) velocity are the data obtained in the experiments.
		}
		\label{S1}
	\end{center}
\end{figure}

In this section, we show how to classify the motion of a camphor grain observed in the experiments as oscillatory or rest state.
First, we obtained the time series of the position of a camphor grain from the videos.
By taking the difference in time, we also obtained the time series of the velocity of the camphor grain.
We then examined the maximum and minimum values of the position and velocity within each time series of 5~s, which was longer than the period of oscillation (1-2 s), as shown in Fig.~\ref{S1}.
Thus, we obtained 60 data for each trial (video for 300~s).
We calculated the average and standard deviation over more than four trials for each channel length.
We defined the averaged values as amplitudes, which are plotted in Fig.~\ref{exp_amp_vamp}.
It is noted that this method is suitable since it can be applied for both the oscillatory and rest states.

\section{Nondimensionalization of the variables in the model\label{App2}}

First, the evolutional equation for the concentration field, Eq.~\eqref{d.eq.}, is nondimensionalized. 
Hereafter, dimensionless variables are denoted by tildes ($\,\tilde{}\,$).
The dimensionless time, $\tilde{t}$, length, $\tilde{x}$, and concentration fields, $\tilde{c}$, are set as $\tilde{t} = \alpha t$, $\tilde{x} = \sqrt{\alpha/D} x$, and $\tilde{c} (\tilde{x},\tilde{t}) = c (x,t) / c_0$, respectively.
By substituting these dimensionless variables into Eq.~\eqref{d.eq.}, we obtain
\begin{equation}
\label{A1}\frac{\partial \tilde{c}(\tilde{x},\tilde{t})}{\partial \tilde{t}} = \frac{\partial ^2 \tilde{c}(\tilde{x},\tilde{t})}{\partial \tilde{x}^2} - \tilde{c}(\tilde{x},\tilde{t}) + \frac{1}{c_0 \alpha} f \left (\sqrt{\frac{D}{\alpha}}\tilde{x}; X \left ( \frac{\tilde{t}}{\alpha} \right )  \right ).
\end{equation}
Here, the source term in Eq.~\eqref{A1}, $f$, is rewritten as
\begin{align}
\tilde{f}(\tilde{x}; \tilde{X}(\tilde{t})) &= \frac{1}{c_0 \alpha} f \left ( \sqrt{\frac{D}{\alpha}}\tilde{x}; X \left ( \frac{\tilde{t}}{\alpha} \right )  \right ) \nonumber \\ &= \frac{1}{\sqrt{\alpha D}} \delta \left ( \tilde{x} - \tilde{X} \left ( \tilde{t} \right ) \right ),
\end{align}
where $\tilde{X}(\tilde{t}) = \sqrt{\alpha / D} X(\tilde{t} / \alpha)$.
Here, we use $\delta(ax) = \delta(x) / |a|$.
We then obtain the dimensionless equation for the concentration field as
\begin{align}
\frac{\partial \tilde{c}(\tilde{x},\tilde{t})}{\partial \tilde{t}} = \frac{\partial ^2 \tilde{c}(\tilde{x},\tilde{t})}{\partial \tilde{x} ^2} - \tilde{c}(\tilde{x},\tilde{t}) + \tilde{f} \left (\tilde{x}; \tilde{X}(\tilde{t})\right ),
\end{align}
where
\begin{equation}
\tilde{f}(\tilde{x}; \tilde{X}(\tilde{t})) = \frac{1}{\sqrt{\alpha D}} \delta \left ( \tilde{x} - \tilde{X}(\tilde{t}) \right ) .\label{n.d.eq.}
\end{equation}
Next, we demonstrate the nondimensionalization on the equation of motion, Eq.~\eqref{eq.of.m.}. 
The driving force is represented as
\begin{align}
&F \left ( \sqrt{\frac{D}{\alpha}} \tilde{X}(\tilde{t}) c_0 \tilde{c} \left ( \tilde{x}, \tilde{t} \right ) \right ) \nonumber \\
&= - K c_0 \sqrt{\frac{\alpha}{D}} \left ( \left . \frac{\partial \tilde{c} (\tilde{x},\tilde{t})}{\partial \tilde {x}} \right | _{\tilde {x} = \tilde{X}(\tilde{t}) + 0} + \left . \frac{\partial \tilde{c} (\tilde{x},\tilde{t})}{\partial \tilde {x}} \right | _{\tilde {x} = \tilde{X}(\tilde{t}) - 0} \right ). 
\end{align}
The variables, $t$, $x$, $X$, $c$, and $F$, in the equation of motion are replaced by $\tilde{t}$, $\tilde{x}$, $\tilde{X}$, $\tilde{c}$, and $\tilde{F}$, respectively. We then obtain 
\begin{equation}
\tilde{m} \frac{d^2 \tilde{X}(\tilde{t})}{d \tilde{t}^2} = - \tilde{\eta} \frac{d \tilde{X}(\tilde{t})}{d \tilde{t}} + \tilde{F} \left ( \tilde{X} (\tilde{t}) ; \tilde{c} \left ( \tilde{x}, \tilde{t} \right ) \right ),
\end{equation}
where
\begin{align}
& \tilde{F}(\tilde{X}(\tilde{t}) ; \tilde{c} \left ( \tilde{x}, \tilde{t} \right ) ) \nonumber\\
& = - \left ( \left . \frac{\partial \tilde{c} (\tilde{x},\tilde{t})}{\partial \tilde {x}} \right |_{\tilde {x} = \tilde{X}(\tilde{t}) + 0} + \left . \frac{\partial \tilde{c} (\tilde{x},\tilde{t})}{\partial \tilde {x}} \right | _{\tilde {x} = \tilde{X}(\tilde{t}) - 0} \right ) \label{n.eq.of.m.}.
\end{align}
Here, we define $\tilde{m} = m \alpha D / (K c_0)$ and $\tilde{\eta} = \eta D / (K c_0)$.

\section{Detailed calculation of the reduction of the mathematical model\label{App3}}

In this section, we show the detailed calculation of the reduction of the mathematical model.
By using the partial integration repetitively, Eq.~\eqref{integralform} is expanded as
\begin{align}
c_k =& 2 e^{-At} \int_{-\infty}^{t} d{t}' \cos \left( \kappa X({t}') \right ) e^{A{t}'} \nonumber \\
=& \frac{2 e^{-At}}{A} \cos \left( \kappa X(t) \right ) e^{At} \nonumber \\
&- \frac{2 e^{-At}}{A} \int_{-\infty}^{t} d{t}' \left ( -\kappa \dot{X}({t}') \sin \left( \kappa X({t}') \right ) \right ) e^{A{t}'} \nonumber \\
=& \frac{2 e^{-At}}{A} \cos \left( \kappa X(t) \right ) e^{At} \nonumber \\
&+ \frac{2 \kappa e^{-At}}{A^2} \dot{X}(t) \sin \left( \kappa X(t) \right ) e^{A t} \nonumber \\
& - \frac{2 \kappa e^{-At}}{A^2} \int_{-\infty}^{t} d{t}' \left \{ \ddot{X}({t}') \sin \left( \kappa X({t}') \right ) \right. \nonumber \\ & \left.
\qquad\qquad + \kappa \dot{X}({t}') ^2 \cos \left( \kappa X({t}') \right ) \right \} e^{A{t}'},
\end{align}
where $\kappa = \kappa(k) = k \pi / R$, and $A = A(k) = k^2 \pi^2 / R^2 + 1$ and the dots denote the time derivative~\cite{Ohta2009}.
We obtain
\begin{align}
c_k=&\frac{1}{A} \cos \left( \kappa X(t) \right ) + \frac{\kappa}{A^2} \dot{X} \sin \left( \kappa X(t) \right ) \nonumber \\
& - \frac{\kappa}{A^3} \left \{ \ddot{X}(t) \sin \left( \kappa X(t) \right ) + \kappa \dot{X}(t)^2 \cos \left( \kappa X(t) \right ) \right \} \nonumber \\
& - \frac{\kappa^3}{A^4} \dot{X}(t)^3 \sin \left( \kappa X(t) \right ) \nonumber \\
& + \mathrm{(higher \; order \; terms \; \& \; cross \; terms)}.
\end{align}
It is noted that the higher-order and cross terms, such as $\dddot{X}$, $\dot{X}^4$, and $\dot{X}^2 \ddot{X}$, are neglected.

We then calculate the driving force.
By using the expanded form of the concentration field in the Fourier space, the gradient of the concentration field is expressed as
\begin{widetext}
\begin{eqnarray}
\begin{split}
\frac{\partial c}{\partial x}
=& - \frac{1}{R} \sum _{k=1}^{\infty} \left ( \frac{2 \kappa(k)}{A(k)} \cos \left( \kappa(k) X(t) \right ) \sin \left( \kappa(k) x \right ) \right ) - \frac{1}{R} \sum _{k=1}^{\infty} \left ( \frac{2 \kappa(k)^2}{A(k)^2} \sin \left( \kappa(k) X(t) \right ) \sin \left( \kappa(k) x \right ) \right ) \dot{X}(t) \\
& + \frac{1}{R} \sum _{k=1}^{\infty} \left ( \frac{2 \kappa(k)^2}{A(k)^3} \sin \left( \kappa(k) X(t) \right ) \sin \left( \kappa(k) x \right ) \right ) \ddot{X}(t) + \frac{1}{R} \sum _{k=1}^{\infty} \left ( \frac{2 \kappa(k)^3}{A(k)^3} \cos \left( \kappa(k) X(t) \right ) \sin \left( \kappa(k) x \right )  \right ) \dot{X}(t)^2 \\
& + \frac{1}{R} \sum _{k=1}^{\infty} \left ( \frac{2 \kappa(k)^4}{A(k)^4} \sin \left( \kappa(k) X(t) \right ) \sin \left( \kappa(k) x \right ) \right ) \dot{X}(t)^3.
\end{split}
\end{eqnarray}
By taking the infinite summations and substituting them into the definition of the driving force in Eq.~\eqref{nd_d.f.}, we obtain
\begin{eqnarray}
\begin{split}
F =& \frac{\sinh (R -2 X(t))}{\sinh R} \\
& + \left [ \frac{\cosh R}{2 \sinh R} - \frac{R}{2 \sinh^2 R} - \frac{\cosh (R - 2 X(t))}{2 \sinh R}\right . \\
& \qquad \qquad  \left . - \frac{1}{2} \left ( \frac{(R - 2 X(t)) \sinh (R - 2 X(t))}{\sinh R} - \frac{R \cosh (R - 2 X(t)) \cosh R}{\sinh^2 R} \right ) \right ] \dot{X}(t) \\
& + \left [ - \frac{\cosh R}{8 \sinh R} - \frac{R}{8 \sinh^2 R} + \frac{R^2 \cosh R}{4 \sinh^3 R} + \frac{\cosh (R - 2 X(t))}{8 \sinh R} \right . \\
& \qquad -  \frac{1}{8} \left ( \frac{(R - 2 X(t)) \sinh (R - 2 X(t))}{\sinh R} - \frac{R \cosh (R - 2 X(t)) \cosh R}{\sinh^2 R} \right ) \\
& \qquad - \frac{1}{8} \left ( \frac{(R - 2 X(t))^2 \cosh (R - 2 X(t) )}{\sinh R} - 2 \frac{R (R - 2 X(t)) \sinh (R - 2 X(t)) \cosh R}{\sinh^2 R} \right . \\
& \qquad \qquad \qquad \left . \left . - \frac{R^2 \cosh (R -2 X(t))}{\sinh R} + 2 \frac{R^2 \cosh (R - 2 X(t)) \cosh^2 R}{\sinh^3 R} \right ) \right ] \ddot{X}(t) \\
& + \left [ \frac{3}{8} \left ( \frac{(R - 2 X(t)) \cosh (R - 2 X(t))}{\sinh R} - \frac{R \sinh (R - 2 X(t)) \cosh R}{\sinh^2 R}\right ) \right . \\
& \qquad + \frac{1}{8} \left ( \frac{(R - 2 X(t))^2 \sinh (R - 2 X(t))}{\sinh R} - 2 \frac{R (R - 2 X(t)) \cosh (R - 2 X(t)) \cosh R}{\sinh^2 R} \right . \\
& \qquad \qquad \qquad \left . \left . - \frac{R^2 \sinh (R - 2 X(t)) }{\sinh R} + 2 \frac{R^2 \sinh (R - 2 X(t)) \cosh^2 R}{\sinh^3 R} \right ) \right ] \dot{X}(t)^2 \\
& + \left [- \frac{\cosh R}{16\sinh R} - \frac{R}{16\sinh^2 R} +\frac{R^2 \cosh R}{4 \sinh^3 R} + \frac{R^3}{24 \sinh^2 R} - \frac{R^3 \cosh^2 R}{8 \sinh^4 R} \right . \\
& \qquad + \frac{1}{16} \frac{\cosh (R - 2 X(t))}{\sinh R} - \frac{1}{16} \left ( \frac{(R - 2 X(t)) \sinh (R - 2X(t))}{\sinh R} - \frac{R \cosh (R - 2 X(t)) \cosh R}{\sinh^2 R} \right ) \\
& \qquad - \frac{1}{8} \left ( \frac{(R - 2 X(t))^2 \cosh (R - 2 X(t))}{\sinh R} -2 \frac{R (R - 2 X(t)) \sinh (R - 2 X(t)) \cosh R}{\sinh^2 R} \right . \\
& \qquad \qquad \left . - \frac{R^2 \cosh (R - 2 X(t))}{\sinh R} + 2 \frac{R^2 \cosh (R - 2 X(t)) \cosh^2 R}{\sinh^3 R}\right ) \\
& \qquad - \frac{1}{48} \left ( \frac{(R - 2 X(t))^3 \sinh (R - 2 X(t))}{\sinh R} -3 \frac{R (R - 2 X(t))^2 \cosh (R - 2 X(t)) \cosh R}{\sinh^2 R} \right . \\
& \qquad \qquad -3 \frac{R^2 (R - 2 X(t)) \sinh (R - 2 X(t))}{\sinh R} + 6 \frac{R^2 (R - 2 X(t)) \sinh (R - 2 X(t)) \cosh^2 R}{\sinh^3 R} \\
& \qquad \qquad \qquad \left . \left . + 5 \frac{R^3 \cosh (R - 2 X(t)) \cosh R}{\sinh^2 R} - 6 \frac{R^3 \cosh (R - 2 X(t)) \cosh^3 R}{\sinh^4 R} \right ) \right ] \dot{X}(t)^3.
\end{split}
\label{r.df}
\end{eqnarray}
\end{widetext}
We expand Eq.~\eqref{r.df} around $X=R/2$, and obtain Eq.~\eqref{df}.

\section{Peak of the bifurcation curve\label{App4}}

As shown in Figs.~\ref{pd_ac} and \ref{pd_nc}, the bifurcation curve $C(R)$ has a peak around $R = 2$. 
This peak indicates that a camphor grain can continue moving with the largest resistance coefficient in the water channel with the length of $R \sim 2$.
To explain the existence of this peak, we consider the situation that there is only one boundary for simplicity.

\begin{figure}
	\begin{center}
		\includegraphics{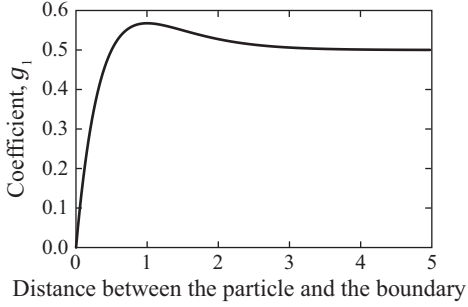}
		\caption{Schematic illustration of the situation. A camphor grain is approaching the boundary with a constant velocity $v$.}
		\label{schematic_illustration}
	\end{center}
\end{figure}

\begin{figure}
	\begin{center}
		\includegraphics{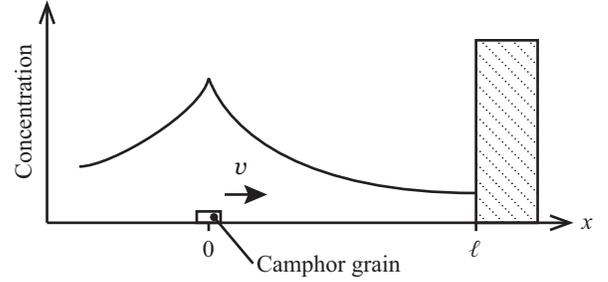}
		\caption{Dependence of the coefficient of the first order of the velocity, $g_1(\ell)$, on the distance between the camphor grain and the boundary, $\ell$.
				$\ell$ is a distance between the camphor grain and the boundary.}
		\label{kabe}
	\end{center}
\end{figure}

When a camphor grain is moving with a constant velocity $v$ and is located at the origin $x=0$ as shown in Fig.~\ref{schematic_illustration}, the concentration field of the camphor molecules is given by
\begin{align}
\label{sol.1boundary}
c(x;v) = 
\left \{
\begin{array}{ll}
\displaystyle{ \frac{1} {\sqrt{v^2+4}} e^{\lambda_{+} x} + \frac{1} {\sqrt{v^2+4}} e^{\lambda_{-} (-x + 2\ell)} }, (x<0),\\
\displaystyle{ \frac{1} {\sqrt{v^2+4}} e^{\lambda_{-} x} + \frac{1} {\sqrt{v^2+4}} e^{\lambda_{-} (-x + 2\ell)} }, (x>0),
\end{array}
\right.
\end{align}
where $\lambda_{\pm} = -v/2 \pm \sqrt{v^2/4+1}$ and the boundary is located at $x=\ell$.
The profile in Eq.~\eqref{sol.1boundary} was obtained by solving Eq.~\eqref{nd_d.eq.} with the Neumann condition at the boundary.
We calculated the driving force working on the camphor grain using Eqs.~\eqref{sol.1boundary} and \eqref{nd_d.f.}, and obtained
\begin{align}
F(v,\ell) = g_0(\ell) + g_1(\ell) v + \mathcal{O}(v^2),
\end{align}
where $\displaystyle{g_0(\ell) = -e^{-2\ell}}$, $\displaystyle{g_1(\ell) = 1/2 + (\ell - 1/2) e^{-2\ell}}$.
The coefficient $g_1(\ell)$ is plotted against $\ell$ in Fig.~\ref{kabe}.
$g_1(\ell)$ has a peak around $\ell=1$.
This peak means that the velocity-dependent component of the driving force takes the largest absolute value when the distance between the camphor grain and the boundary, $\ell$, is around 1.
The bifurcation curve $C(R)$ is the coefficient of the first order of velocity in the expression of the driving force (Eq.~\eqref{df}) in the case where a camphor grain is in a finite region with two boundaries.
In the finite system, the distance between the center position and the boundaries of the system is 1 for $R=2$, i.e., the peak of $g_1(\ell)$ corresponds to that of $C(R)$.
Thus, the reason why $C(R)$ has the peak around $R \sim 2$ can be understood by considering the boundary effect.

\section{supercritical or subcritical Hopf bifurcation\label{App5}}

\begin{figure}
	\begin{center}
		\includegraphics{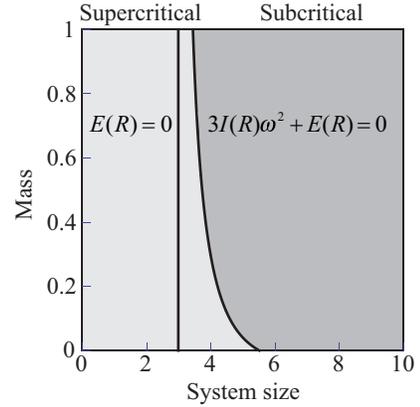}
		\caption{Plot of the border between the supercritical and subcritical Hopf bifurcations, $3I(R)\left[\omega(R,m)\right]^2+E(R)=0$, on the system size and mass ($R-m$) plane.
		For smaller and larger system sizes, supercritical and subcritical Hopf bifurcations occur, respectively.
		It is noted that the curve $3I(R) \left[\omega(R,m)\right]^2+E(R)=0$ converges to $E(R)=0$ since $\omega(R,m)$ goes to 0 when the mass $m$ goes to infinity.}
		\label{super_sub}
	\end{center}
\end{figure}
Hopf bifurcations are classified into supercritical and subcritical using a criterion for the two-dimensional dynamical system~\cite{Guckenheimer}.
By applying the criterion, a supercritical Hopf bifurcation occurs for $3I(R) \left[\omega(R,m)\right]^2+E(R) > 0$ and a subcritical Hopf bifurcation occurs for $3I(R)\left[\omega(R,m)\right]^2+E(R) < 0$ in the reduced equation~\eqref{rdc.eq}.
Here we defined $\omega(R,m) = \sqrt{-A(R)/(m-G(R))}$.
$E(R)$, $I(R)$, and $\omega(R,m)$ depend on the system size, $R$, and $\omega(R,m)$ also depends on the mass, $m$. 
The border between the supercritical and subcritical Hopf bifurcations, $3I(R) \left[\omega(R,m)\right]^2+E(R)=0$, is shown in Fig.~\ref{super_sub}.
\begin{figure}
	\begin{center}
		\includegraphics{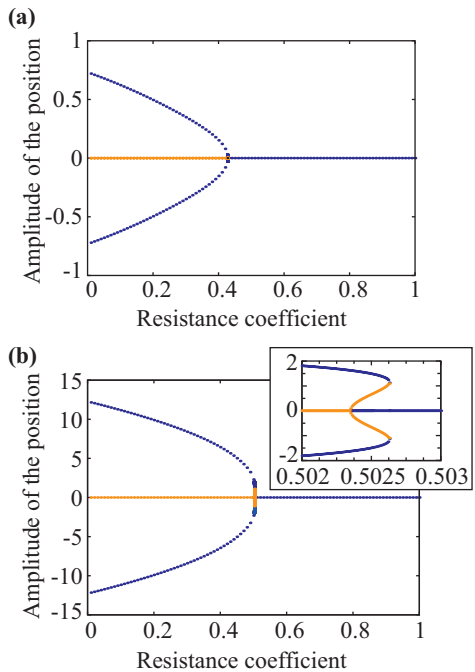}
		\caption{Bifurcation diagrams depending on the resistance coefficient $\eta$ for the system size (a) $R=1$ and (b) $R=8$.
		The stable and unstable amplitudes are plotted with blue and orange dots.
		The dots located at zero amplitude correspond to the rest state at the center position of the system.
		In the inset of (b), the magnified plot around the bifurcation point is shown.
		The mass $m$ is set to $0.01$.
		}
		\label{bif_str}
	\end{center}
\end{figure}

To confirm the analytical results for the type of Hopf bifurcation, i.e., supercritical or subcritical, the stable and unstable branches of the amplitude depending on the resistance coefficient were numerically obtained based on Eqs.~\eqref{rdc.eq} and \eqref{df}.
The stable and unstable amplitudes were obtained by numerical calculations performed in forward and backward time directions, respectively.
The bifurcation diagrams for the system sizes $R = 1$ and $8$ are shown in Fig.~\ref{bif_str}.
The stable and unstable oscillations were seen around the bifurcation points for $R=1$ and $R=8$, respectively, which is consistent with the analytical results.
For the subcritical Hopf bifurcation, the regime where the unstable oscillation existed was narrow, and the stable oscillation with finite amplitude seems to appear just above the bifurcation point of resistance coefficient as shown in Fig.~\ref{bif_str}(b).

\end{document}